\documentclass[11pt,a4paper]{article}
\pdfoutput=1   
\usepackage[hyperref]{acl2020}
\usepackage{times}
\usepackage{latexsym}

\newcommand{\egcvalue}[1]{\textbf{\textit{#1}}}
\usepackage{enumitem}
\usepackage{color}
\usepackage{tcolorbox}
\usepackage{tikz}
\usepackage{amssymb}

\usepackage{microtype}

\aclfinalcopy 

\definecolor{azure}{rgb}{0.0, 0.5, 1.0}
\tcbuselibrary{skins}
\tcbset{enhanced}

\makeatletter
\newcommand\footnoteref[1]{\protected@xdef\@thefnmark{\ref{#1}}\@footnotemark}
\makeatother

\usepackage{color}
\usepackage{tcolorbox}
\usepackage{tikz}
\usepackage{amssymb}

\definecolor{azure}{rgb}{0.0, 0.5, 1.0}
\definecolor{question-colour}{HTML}{e7f1ff}
\definecolor{section-colour}{HTML}{9ec6ff}
\definecolor{border-colour}{HTML}{0065f3}

\tcbuselibrary{skins}
\tcbset{enhanced}

\newcommand{\qsecboxqtn}[1]{\begin{tcolorbox}[colback=question-colour,colframe=border-colour]\raggedright{#1}\end{tcolorbox}}

\usepackage{wasysym}
\newcommand{\radiobutton}[0]{\ooalign{\hidewidth\cr$\ocircle$}}
\newcommand{\checkbox}[0]{\ooalign{\hidewidth\cr$\square$}}

\title{HEDS 3.0: The Human Evaluation Data Sheet Version 3.0}

\author{Anya Belz \and Craig Thomson\\
  ADAPT, Dublin City University\\
  Dublin, Ireland \\
  \texttt{\{anya.belz,craig.thomson\}@dcu.ie}
}

\date{}

\begin{document}
\maketitle
\begin{abstract}
This paper presents version 3.0 of the Human Evaluation Datasheet (HEDS). This update is the result of our experience using HEDS in the context of numerous recent human evaluation experiments, including reproduction studies, and of feedback received. Our main overall goal was to improve clarity, and to enable users to complete the datasheet more consistently and comparably. 
The HEDS 3.0 package consists of the digital data sheet, documentation, and code for exporting completed data sheets as latex files, all available from the HEDS GitHub.\footnote{\url{https://github.com/DCU-NLG/HEDS-3.0}}
\end{abstract}


\section{Introduction}

We present the third version of the Human Evaluation Datasheet (HEDS~3.0), the template for describing single human evaluation experiments. Like its predecessors, HEDS~1.0 \cite{shimorina:belz:2021} and HEDS~2.0 \cite{shimorina:belz:2022}, it is intended to be generally applicable to human evaluations across NLP, uses multiple-choice questions where possible, for ease of comparability across experiments, and is divided into five sections as follows:
\begin{enumerate}\itemsep=0pt
    \item Main Reference and Supplementary Resources (Questions 1.1.1--1.3.2.3);
    \item Evaluated System(s) (Questions 2.1--2.5);
    \item Sample of System Outputs, Evaluators and Experimental Design (Questions 3.1.1--3.3.8);
    \item Definition and Operationalisation of Quality Criteria (Questions 4.1.1--4.3.12.2); 
    \item Ethics (Questions 5.1--5.4).
\end{enumerate}

\noindent A summary of changes since HEDS~2.0 can be found in Appendix~\ref{appsec:delta}.

\section{Credits}

Questions 2.1--2.5 relating to evaluated system(s), and 4.3.1--4.3.8 relating to response elicitation, are based on \citet{howcroft-etal-2020-twenty}, with some significant changes. Questions 4.1.1--4.2.3 relating to quality criteria, and some of the questions about system outputs, evaluators, and experimental design (3.1.1--3.2.3, 4.3.5, 4.3.6, 4.3.9--4.3.11) are based on \citet{belz-etal-2020-disentangling}. HEDS was also informed by \citet{van2019best} and \citet{vanderlee2021}, and by \cite{gehrmann2021gem}'s data card guide. More generally, the original inspiration for creating a ‘datasheet’ for describing human evaluation experiments of course comes from seminal papers by \citet{bender-friedman-2018-data}, \citet{mitchell2019modelcards}, and \citet{gebru2018datasheets}.

\section{Paper Overview}

The paper is structured as follows. Section~\ref{sec:resources} lists the components that make up the HEDS 3.0 package. Section~\ref{sec:pres-conventions} has notes on how we present the questions in the HEDS form in this paper, and what the different question (and answer) types are.

Section~\ref{sec:heds-instructions} presents the parts of the instructions from the HEDS form that relate to the content of the form (rather than its interactive completion online), and Section~\ref{sec:heds-form} presents the HEDS form itself, verbatim. The latter of these two sections was automatically generated from the form.
 
Finally, in Section~\ref{sec:add-comments} we provide additional explanation for some aspects that we know from experience users may find more difficult.

\section{Package Components}\label{sec:resources}

The HEDS 3.0 package consists of the following three resources, all available at \url{https://github.com/DCU-NLG/HEDS-3.0}:

\begin{enumerate}\itemsep=0pt
    \item The HEDS 3.0 form and instructions for completion: available for online completion at \url{https://nlp-heds.github.io};
    \item Description and completion guidance: this document;
    \item Scripts for exporting completed HEDS 3.0 forms to alternative formats, including Latex.
\end{enumerate}

\section{Question Types and Presentation}\label{sec:pres-conventions}

In Section~\ref{sec:heds-form} we present the HEDS form in its entirety, in a similar look/feel to the online version that users complete (in fact, the whole section is generated automatically from the form).

Questions come in the following types and presentation formats: 

\begin{enumerate}\itemsep=0pt
    \item Multiple-choice list, select one: radio buttons.
    \item Multiple-choice list, select all that apply: check boxes.
    \item Integer field: entry is validated to be exactly one integer.
    \item Short text box, enter one type of information (a URL, a value range, etc.).
    \item Longer text box: enter (a) more comprehensive information, and/or (b) information that depends on given factors.
\end{enumerate}

\section{HEDS Instructions}\label{sec:heds-instructions}

This is the Human Evaluation Datasheet (HEDS) form which is designed to record full details of human evaluation experiments in Natural Language Processing (NLP), addressing a history of details often going unreported in the field (in extreme cases, no details at all are reported). Reporting such details is crucial for gauging the reliability of results, determining comparability with other experiments, and for assessing reproducibility \citep{belz-etal-2023-missing,belz-etal-2023-non, thomson-etal-2024-common, thomson-belz-2024-mostly-automatic}. Having a standard set of questions to answer (as provided by HEDS) means not having to worry about what information to include or in what detail, as well as the information being in a format directly comparable to information reported for other human evaluation experiments. To maximise standardisation, questions are in multiple-choice format where possible.

The HEDS form is divided into five main sections, containing questions that record information about resources, evaluated system(s), test set sampling, quality criteria assessed, and ethics, respectively. Within each of the main sections there can be multiple subsections which can be expanded or collapsed.

Each HEDS question comes with instructions and notes to help with answering it, except where the task is exceedingly simple (e.g.\ when a contact email address is asked for).

HEDS Section 4 needs to be completed for each quality criterion that is evaluated in the experiment. Instructions on how to do this are shown at the start of HEDS Section 4.

The form is not submitted to any server when it is completed, and instead needs to be downloaded to a local file. A tool is available in the GitHub repository for converting the file to latex format (which we used to generate the next section).



We recognise that completing a form of this length and level of detail constitutes an overhead in terms of time and effort, especially the first time a HEDS form is completed when the learning curve is steepest. However, this overhead does go down substantially with each use of HEDS, and, we believe, is far outweighed by the benefits: increased scientific rigour, reliability and repeatability.

We envisage the main uses of HEDS to be as follows. Ideally, it should be completed before a human evaluation experiment is run, at the point when the design is final, as part of a formal preregistration process. Once the experiment has been run, the information in the sheet can be updated if necessary, e.g.\ if the final number of evaluators had to change due to unforeseen circumstances.

Another use is for the purpose of reporting the details of a completed experiment. For this, the completed HEDS sheet can be automatically converted to Latex, ready for inclusion in the supplementary material.

A third use is for carrying out reproducibility studies, as has been done extensively in the ReproGen and ReproNLP shared tasks \cite{belz20222022,belz20242024}. Here, the HEDS sheets were used to ensure that original work and reproduction experiment had the same properties, hence can be expected to produce similar results.

\section{HEDS Form in its Entirety}\label{sec:heds-form}

\section*{HEDS Section~1:~Main Reference and Supplementary Resources}

\subsection*{1.1$\;$~Main reference}

\vspace{0.2cm}

\subsection*{\qsecboxqtn{Question 1.1.1: Where can the main reference for the evaluation experiment be found?}}

\vspace{.075cm}
\noindent \textit{Multiple-choice options (select one)}:
\begin{enumerate}[itemsep=0cm,leftmargin=0.5cm]
	\item[\radiobutton] \egcvalue{The main paper reporting the experiment is here (enter URL).} 

	\item[\radiobutton] \egcvalue{An unpublished report describing the experiment can be found here (enter URL).} 

	\item[\radiobutton] \egcvalue{No report describing the experiment is available and this sheet will be uploaded for preregistration here (enter URL).} 

	\item[\radiobutton] \egcvalue{No report describing the experiment is available and no pregistration is not planned.} 
\vspace{-0.4cm}

\end{enumerate} 

\subsection*{\qsecboxqtn{Question 1.1.2: Which experiment is this form being completed for?}}
\vspace{-0.075cm}
\noindent \textit{What to enter in the text box}:~~Referring to the main reference entered for Question 1.1.1, identify the experiment that you're completing this form for (see instructions section at the start for explanation of term `experiment'), in particular to differentiate this experiment from any others that you are carrying out as part of the same overall work: (a) if a link for a published paper was entered under Question 1.1.1, give here the section(s) and/or table(s) that best identify the experiment, plus a brief description for clarity; (b) if `preregistration' or `unpublished' was selected, enter a brief description of the experiment, mentioning quality criteria, dataset and systems.

\subsection*{1.2$\;$~Supplementary resources}
\vspace{.1cm}

\subsection*{\qsecboxqtn{Question 1.2: Where can the resources that were used in the evaluation experiment be found?}}

\noindent \textit{Multiple-choice options (select one)}:
\begin{enumerate}[itemsep=0cm,leftmargin=0.5cm]
	\item[\radiobutton] \egcvalue{The resources used in the experiment can be found here (enter URL(s)).} 

	\item[\radiobutton] \egcvalue{No resources shared.} 

\end{enumerate}\vspace{0.2cm}

\subsection*{1.3$\;$~Contact Details}
\vspace{.1cm}

\subsection*{1.3.1$\;$~Details of the person completing this sheet.}
\vspace{.1cm}

\subsection*{\qsecboxqtn{Question 1.3.1.1: Name of the person completing this sheet.}}

\subsection*{\qsecboxqtn{Question 1.3.1.2: Affiliation of the person completing this sheet.}}

\subsection*{\qsecboxqtn{Question 1.3.1.3: Email address of the person completing this sheet.}}
\vspace{.3cm}

\subsection*{1.3.2$\;$~Details of the contact author}
\vspace{.1cm}

\subsection*{\qsecboxqtn{Question 1.3.2.1: Name of the contact author.}}

\subsection*{\qsecboxqtn{Question 1.3.2.2: Affiliation of the contact author.}}

\subsection*{\qsecboxqtn{Question 1.3.2.3: Email address of the contact author.}}

\vspace{.8cm}

\section*{HEDS Section~2:~Evaluated System(s)}

\noindent \textit{Notes:} Questions 2.1--2.5 in this section record information about the system(s) that are evaluated in the experiment  this sheet is being completed for. The input, output and task questions  are closely interrelated: the answer to one partially determines the answer to the others, as indicated for some combinations of answers under Question 2.3.

\vspace{-0.5cm}

\subsection*{\qsecboxqtn{Question 2.1: What type of input do the evaluated system(s) take?}}

\vspace{-.1cm}

\noindent \textit{Notes:} The term `input' here refers to the text, representations and/or data structures that all of the evaluated systems take as input (including prompts).  This question is about input \textit{type}, regardless of number. E.g.\  if the input is a set of documents, you would still select `text: document' below.

\vspace{.1cm}
\noindent \textit{Check-box options (select all that apply)}:
\begin{enumerate}[itemsep=0cm,leftmargin=0.5cm]
	\item[\checkbox] \egcvalue{Raw/structured data}:~~Numerical, symbolic, and other data, possibly structured into trees, graphs, graphical models, etc. E.g.\  the input to Referring Expression Generation (REG), end-to-end text generation, etc.  NB: excludes linguistic representations. 

	\item[\checkbox] \egcvalue{Deep linguistic representation (DLR)}:~~Any of a variety of deep, underspecified, semantic representations, such as abstract meaning representations (AMRs; \citet{banarescu-etal-2013-abstract}) or discourse representation structures (DRSs; \citet{kamp-reyle-2013}). 

	\item[\checkbox] \egcvalue{Shallow linguistic representation (SLR)}:~~Any of a variety of shallow, syntactic representations, e.g.\  Universal Dependency (UD) structures; typically the input to surface realisation. 

	\item[\checkbox] \egcvalue{Text: subsentential unit of text}:~~Unit(s) of text shorter than a sentence, e.g.\  Referring Expressions (REs), verb phrase, text fragment of any length; includes titles/headlines. 

	\item[\checkbox] \egcvalue{Text: sentence}:~~Single sentence(s). 

	\item[\checkbox] \egcvalue{Text: multiple sentences}:~~Sequence(s) of multiple sentences, without any document structure. 

	\item[\checkbox] \egcvalue{Text: document}:~~Text(s) with document structure, such as a title, paragraph breaks or sections, e.g.\  a set of news reports for summarisation. 

	\item[\checkbox] \egcvalue{Text: dialogue}:~~Dialogue(s) of any length, excluding a single turn which would come under one of the other text types. 

	\item[\checkbox] \egcvalue{Text: other (please describe)}:~~Input is text but doesn't match any of the above text categories. 

	\item[\checkbox] \egcvalue{Speech}:~~Recording(s) of speech. 

	\item[\checkbox] \egcvalue{Visual}:~~Image(s) or video(s). 

	\item[\checkbox] \egcvalue{Multi-modal}:~~Select this option if input is\textit{always} a combination of multiple modalities. Also select other options in this list to different elements of the multi-modal input. 

	\item[\checkbox] \egcvalue{Control feature}:~~Feature(s) or parameter(s) specifically present to control a property of the output text, e.g.\  positive stance, formality, author style. 

	\item[\checkbox] \egcvalue{No input (please explain)}:~~If there are no system inputs, select this option and explain why. 

	\item[\checkbox] \egcvalue{Other (please describe)}:~~If input is none of the above, select this option and describe it. 

\end{enumerate}

\subsection*{\vspace{-0.675cm}\qsecboxqtn{Question 2.2: What type of output do the evaluated system(s) generate?}}

\noindent \textit{Notes:} The term `output' here refers to the text, representations and/or data structures that all of the evaluated systems produce as output. This question is about output type, regardless of number. E.g.\  if the output is a set of documents, you would still select `text: document' below.

\vspace{.1cm}
\noindent \textit{Check-box options (select all that apply)}:
\begin{enumerate}[itemsep=0.1cm,leftmargin=0.5cm]
	\item[\checkbox] \egcvalue{Raw/structured data}:~~Numerical, symbolic, and other data, possibly structured into trees, graphs, graphical models, etc. E.g.\  the input to Referring Expression Generation (REG), end-to-end text generation, etc.  NB: excludes linguistic representations. 

	\item[\checkbox] \egcvalue{Deep linguistic representation (DLR)}:~~Any of a variety of deep, underspecified, semantic representations, such as abstract meaning representations (AMRs; \citet{banarescu-etal-2013-abstract}) or discourse representation structures (DRSs; \citet{kamp-reyle-2013}). 

	\item[\checkbox] \egcvalue{Shallow linguistic representation (SLR)}:~~Any of a variety of shallow, syntactic representations, e.g.\  Universal Dependency (UD) structures; typically the input to surface realisation. 

	\item[\checkbox] \egcvalue{Text: subsentential unit of text}:~~Unit(s) of text shorter than a sentence, e.g.\  Referring Expressions (REs), verb phrase, text fragment of any length; includes titles/headlines. 

	\item[\checkbox] \egcvalue{Text: sentence}:~~Single sentence(s). 

	\item[\checkbox] \egcvalue{Text: multiple sentences}:~~Sequence(s) of multiple sentences, without any document structure. 

	\item[\checkbox] \egcvalue{Text: document}:~~Text(s) with document structure, such as a title, paragraph breaks or sections, e.g.\  a set of news reports for summarisation. 

	\item[\checkbox] \egcvalue{Text: dialogue}:~~Dialogue(s) of any length, excluding a single turn which would come under one of the other text types. 

	\item[\checkbox] \egcvalue{Text: other (please describe)}:~~Input is text but doesn't match any of the above text categories. 

	\item[\checkbox] \egcvalue{Speech}:~~Recording(s) of speech. 

	\item[\checkbox] \egcvalue{Visual}:~~Image(s) or video(s). 

	\item[\checkbox] \egcvalue{Multi-modal}:~~Select this option if input is\textit{always} a combination of multiple modalities. Also select other options in this list to different elements of the multi-modal input. 

	\item[\checkbox] \egcvalue{No input (please explain)}:~~If there are no system inputs, select this option and explain why. 

	\item[\checkbox] \egcvalue{Other (please describe)}:~~If input is none of the above, select this option and describe it. 

\end{enumerate}

\subsection*{\qsecboxqtn{Question 2.3: What is the task that the evaluated system(s) perform in mapping the inputs in Question~2.1 to the outputs in Question~2.2?}}

\noindent \textit{Notes:} This question is about the task(s) performed by the system(s) being evaluated.  This is independent of the application domain (financial reporting, weather forecasting, etc.), or the specific method (rule-based, neural, etc.) implemented in the system. We indicate mutual constraints between inputs, outputs and task for some of the options below.

\vspace{.1cm}
\noindent \textit{Check-box options (select all that apply)}:
\begin{enumerate}[itemsep=0cm,leftmargin=0.5cm]
	\item[\checkbox] \egcvalue{Content selection/determination}:~~Selecting the specific content that will be expressed in the generated text from a representation of possible content. This could be attribute selection for REG (without the surface realisation step). Note that the output here is not text. 

	\item[\checkbox] \egcvalue{Content ordering/structuring}:~~Assigning an order and/or structure to content to be included in generated text. Note that the output here is not text. 

	\item[\checkbox] \egcvalue{Aggregation}:~~Converting inputs (typically \textit{deep linguistic representations} or \textit{shallow linguistic representations}) in some way in order to reduce redundancy (e.g.\  representations for `they like swimming', `they like running' → representation for `they like swimming and running'). 

	\item[\checkbox] \egcvalue{Referring expression generation}:~~Generating \textit{text} to refer to a given referent, typically represented in the input as a set of attributes or a linguistic representation. 

	\item[\checkbox] \egcvalue{Lexicalisation}:~~Associating (parts of) an input representation with specific lexical items to be used in their realisation. 

	\item[\checkbox] \egcvalue{Deep generation}:~~One-step text generation from \textit{raw/structured data} or \textit{deep linguistic representations}. One-step means that no intermediate representations are passed from one independently run module to another. 

	\item[\checkbox] \egcvalue{Surface realisation (SLR to text)}:~~One-step text generation from \textit{shallow linguistic representations}. One-step means that no intermediate representations are passed from one independently run module to another. 

	\item[\checkbox] \egcvalue{Feature-controlled text generation}:~~Generation of text that varies along specific dimensions where the variation is controlled via \textit{control features} specified as part of the input. Input is a non-textual representation (for feature-controlled text-to-text generation select the matching text-to-text task). 

	\item[\checkbox] \egcvalue{Data-to-text generation}:~~Generation from \textit{raw/structured data} which may or may not include some amount of content selection as part of the generation process. Output is likely to be \textit{text:} or \textit{multi-modal}. 

	\item[\checkbox] \egcvalue{Dialogue turn generation}:~~Generating a dialogue turn (can be a greeting or closing) from a representation of dialogue state and/or last turn(s), etc. 

	\item[\checkbox] \egcvalue{Question generation}:~~Generation of questions from given input text and/or knowledge base such that the question can be answered from the input. 

	\item[\checkbox] \egcvalue{Question answering}:~~Input is a question plus optionally a set of reference texts and/or knowledge base, and the output is the answer to the question. 

	\item[\checkbox] \egcvalue{Paraphrasing/lossless simplification}:~~Text-to-text generation where the aim is to preserve the meaning of the input while changing its wording. This can include the aim of changing the text on a given dimension, e.g.\  making it simpler, changing its stance or sentiment, etc., which may be controllable via input features. Note that this task type includes meaning-preserving text simplification (non-meaning preserving simplification comes under \textit{compression/lossy simplification} below). 

	\item[\checkbox] \egcvalue{Compression/lossy simplification}:~~Text-to-text generation that has the aim to generate a shorter, or shorter and simpler, version of the input text. This will normally affect meaning to some extent, but as a side effect, rather than the primary aim, as is the case in \textit{summarisation}. 

	\item[\checkbox] \egcvalue{Machine translation}:~~Translating text in a source language to text in a target language while maximally preserving the meaning. 

	\item[\checkbox] \egcvalue{Summarisation (text-to-text)}:~~Output is an extractive or abstractive summary of the important/relevant/salient content of the input document(s). 

	\item[\checkbox] \egcvalue{End-to-end text generation}:~~Use this option if the system task corresponds to more than one of tasks above, but the system doesn't implement them as separate tasks. 

	\item[\checkbox] \egcvalue{Image/video description}:~~Input includes \textit{visual}, and the output describes it in some way. 

	\item[\checkbox] \egcvalue{Post-editing/correction}:~~The system edits and/or corrects the input text (can itself be the textual output from another system) to yield an improved version of the text. 

	\item[\checkbox] \egcvalue{Other (please describe)}:~~If task is none of the above, Select this option and describe it. 

\end{enumerate}\vspace{0.2cm}

\subsection*{\vspace{-1cm}\qsecboxqtn{Question 2.4: What are the language(s) of the inputs accepted by the system(s)?}}

\vspace{-0.1cm}
\noindent \textit{Notes:} Select any language(s) that apply from this list of standardised full language names as per \href{https://en.wikipedia.org/wiki/List_of_ISO_639-1_codes}{ISO 639-1} (2019).  If language is not (part of) the input, select `N/A'.

\vspace{.1cm}
\noindent \textit{Check-box options (select all that apply)}:
\begin{enumerate}[itemsep=0cm,leftmargin=0.5cm]
	\item[\checkbox] \egcvalue{N/A (please explain)}:~~No language in the input. 

	\item[\checkbox] \egcvalue{Abkhazian}:~~Also known as Abkhaz. 

	\item[\checkbox] \egcvalue{Afar.} 

	\item[\checkbox] \egcvalue{Afrikaans.} 

	\item[\checkbox] \textbf{\ldots}
	\item[\checkbox] \egcvalue{Zhuang, Chuang.} 

	\item[\checkbox] \egcvalue{Zulu.} 

	\item[\checkbox] \egcvalue{Other (please describe)}:~~A language that is not on the above list. 

\end{enumerate}\vspace{0.2cm}

\subsection*{\vspace{-1cm}\qsecboxqtn{Question 2.5: What are the language(s) of the outputs  produced by the system?}}

\vspace{-0.1cm}
\noindent \textit{Notes:} Select any language(s) that apply from this list of standardised full language names as per \href{https://en.wikipedia.org/wiki/List_of_ISO_639-1_codes}{ISO 639-1} (2019).  If language is not (part of) the output, select `N/A'.

\vspace{.1cm}
\noindent \textit{Check-box options (select all that apply)}:
\begin{enumerate}[itemsep=0cm,leftmargin=0.5cm]
	\item[\checkbox] \egcvalue{N/A (please explain)}:~~No language is generated. 

	\item[\checkbox] \egcvalue{Abkhazian}:~~Also known as Abkhaz. 

	\item[\checkbox] \egcvalue{Afar.} 

	\item[\checkbox] \egcvalue{Afrikaans.} 

	\item[\checkbox] \textbf{\ldots}
	\item[\checkbox] \egcvalue{Zhuang, Chuang.} 

	\item[\checkbox] \egcvalue{Zulu.} 

	\item[\checkbox] \egcvalue{Other (please describe)}:~~A language that is not on the above list. 

\end{enumerate}\vspace{0.2cm}

\section*{HEDS Section~3:~Sample of system outputs, evaluators, experimental design}

\subsection*{3.1$\;$~Sample of system outputs (test set)}

\noindent Questions 3.1.1--3.1.3 record information about the size of the sample of outputs (or human-authored stand-ins) evaluated per system, how the sample was selected, and what its statistical power is.

\subsection*{\vspace{-0.7cm}\qsecboxqtn{Question 3.1.1: How many system outputs (or other evaluation items) are evaluated per system?}}
\vspace{-.1cm}

\noindent \textit{What to enter in the text box}:~~The number of system outputs (or other evaluation items) that are evaluated per system by at least one evaluator in the experiment.  For most experiments this should be a single integer.  If the number of outputs varies please explain how and why.

\subsection*{\vspace{-0.6cm}\qsecboxqtn{Question 3.1.2: How are system outputs (or other evaluation items) selected for inclusion?}}

\vspace{-.1cm}
\noindent \textit{Multiple-choice options (select one)}:
\vspace{-.1cm}
\begin{enumerate}[itemsep=0cm,leftmargin=0.5cm]
	\item[\radiobutton] \egcvalue{By simple automatic random selection}:~~Outputs are selected from a larger set by a script using a pseudo-random number generator, without stratification, every-$n$th selection, etc. 

	\item[\radiobutton] \egcvalue{By an automatic random process but using stratified sampling over given properties}:~~Selection is by a random script as above, but with added constraints ensuring that the sample is representative of the set of outputs it is selected from, in terms of given properties, such as sentence length, positive/negative stance, etc. 

	\item[\radiobutton] \egcvalue{By non-random automatic selection}:~~Output sample is selected by a non-randomised automatic process, e.g.\  selecting every $n$th item. 

	\item[\radiobutton] \egcvalue{By manual, arbitrary selection}:~~Output sample was selected by hand, or automatically from a manually compiled list, without specific selection criteria. 

	\item[\radiobutton] \egcvalue{By manual selection aimed at achieving balance or variety relative to given properties}:~~Selection by hand as above, but with specific selection criteria, e.g.\  same number of outputs from each time period. 

	\item[\radiobutton] \egcvalue{Other (please describe)}:~~If selection method is none of the above, select this option and describe it. 

\end{enumerate}\vspace{-0.2cm}

\subsection*{3.1.3$\;$~Statistical power of the sample}

\noindent \textit{Notes:} All evaluation experiments should perform a power analysis to determine an appropriate sample size. If none was performed, enter `N/A' in Questions 3.1.3.1--3.1.3.3

\subsection*{\vspace{-.5cm}
\qsecboxqtn{Question 3.1.3.1: What method of statistical power analysis was used to determine the appropriate sample size?}}

\noindent \textit{What to enter in the text box}:~~The name of the method used, and a URL linking to a reference for the method.

\subsection*{\vspace{-.5cm}\qsecboxqtn{Question 3.1.3.2: What is the statistical power of the sample?}}

\noindent \textit{What to enter in the text box}:~~The numerical results of the statistical power calculation on the output sample obtained with the method in Question 3.1.3.1.

\subsection*{\vspace{-.5cm}\qsecboxqtn{Question 3.1.3.3: Where can other researchers find details of any code used in the power analysis performed?}}

\noindent \textit{What to enter in the text box}:~~A URL linking to any code used in the calculation in Question 3.1.3.2.

\vspace{.1cm}
\subsection*{3.2$\;$~Evaluators}
\vspace{.2cm}

\subsection*{\qsecboxqtn{Question 3.2.1: How many evaluators are there in this experiment?}}
\vspace{-.1cm}

\noindent \textit{What to enter in the text box}:~~A single integer representing the total number of evaluators whose assessments contribute to results in the experiment. Don't count evaluators who performed some evaluations but who were subsequently excluded.
\vspace{.1cm}

\subsection*{3.2.2$\;$~Evaluator Type}
\vspace{.1cm}

\subsection*{\qsecboxqtn{Question 3.2.2.1: Are the evaluators in this experiment domain experts?}}

\vspace{-.1cm}
\noindent \textit{Multiple-choice options (select one)}:
\begin{enumerate}[itemsep=0cm,leftmargin=0.5cm]
	\item[\radiobutton] \egcvalue{Yes}:~~Participants are considered domain experts, e.g.\  meteorologists evaluating a weather forecast generator, or nurses evaluating an ICU report generator. 

	\item[\radiobutton] \egcvalue{No}:~~Participants are not domain experts. 

	\item[\radiobutton] \egcvalue{N/A (please explain).} 

\end{enumerate}

\subsection*{\vspace{-.5cm}\qsecboxqtn{Question 3.2.2.2: Did participants receive any form of payment?}}

\vspace{-.1cm}
\noindent \textit{Multiple-choice options (select one)}:
\begin{enumerate}[itemsep=0cm,leftmargin=0.5cm]
	\item[\radiobutton] \egcvalue{Paid (monetary compensation)}:~~Participants were given some form of monetary compensation for their participation. 

	\item[\radiobutton] \egcvalue{Paid (non-monetary compensation such as course credits)}:~~Participants were given some form of non-monetary compensation for their participation, e.g.\  vouchers, course credits, or reimbursement for travel unless based on receipts. 

	\item[\radiobutton] \egcvalue{Not paid}:~~Participants were not given compensation of any kind (except for receipt-based reimbursement of expenses). 

	\item[\radiobutton] \egcvalue{N/A (please explain).} 

\end{enumerate}

\subsection*{\vspace{-.5cm}\qsecboxqtn{Question 3.2.2.3: Were any of the participants previously known to the authors?}}

\vspace{.1cm}
\noindent \textit{Multiple-choice options (select one)}:
\begin{enumerate}[itemsep=0cm,leftmargin=0.5cm]
	\item[\radiobutton] \egcvalue{Yes}:~~One or more of the researchers running the experiment knew some or all of the participants before recruiting them for the experiment. 

	\item[\radiobutton] \egcvalue{No}:~~None of the researchers running the experiment knew any of the participants before recruiting them for the experiment. 

	\item[\radiobutton] \egcvalue{N/A (please explain).} 

\end{enumerate}\vspace{0.2cm}

\subsection*{\vspace{-.5cm}\qsecboxqtn{Question 3.2.2.4: Were any of the researchers running the experiment among the participants?}}

\vspace{-.15cm}
\noindent \textit{Multiple-choice options (select one)}:
\begin{enumerate}[itemsep=-0.1cm,leftmargin=0.5cm]
	\item[\radiobutton] \egcvalue{Yes}:~~Evaluators include one or more of the researchers running the experiment. 

	\item[\radiobutton] \egcvalue{No}:~~Evaluators do not include any of the researchers running the experiment. 

	\item[\radiobutton] \egcvalue{N/A (please explain).} 

\end{enumerate}\vspace{0.2cm}

\subsection*{\vspace{-1cm}\qsecboxqtn{Question 3.2.3: How are evaluators recruited?}}

\vspace{-.15cm}
\noindent \textit{What to enter in the text box}:~~Explain how your evaluators are recruited. Do you send emails to a given list? Do you post invitations on social media? Posters on university walls? Were there any gatekeepers involved?

\subsection*{\vspace{-.5cm}\qsecboxqtn{Question 3.2.4: What training and/or practice are evaluators given before starting on the evaluation itself?}}

\vspace{-.15cm}
\noindent \textit{What to enter in the text box}:~~Describe any training evaluators were given to prepare them for the evaluation task, including any practice evaluations they did. This includes introductory explanations, e.g.\  on the start page of an online evaluation tool.

\subsection*{\vspace{-.5cm}\qsecboxqtn{Question 3.2.5: What other characteristics do the evaluators have?}}

\vspace{-.15cm}
\noindent \textit{What to enter in the text box}:~~Use this space to list any characteristics not covered in previous questions that the evaluators are known to have, e.g.\  because of information collected during the evaluation. This might include geographic location, educational level, or demographic information such as gender, age, etc. Where characteristics differ among evaluators (e.g.\  gender, age, location etc.), also give numbers for each subgroup.

\subsection*{3.3$\;$~Experimental Design}

\subsection*{\vspace{-.1cm}\qsecboxqtn{Question 3.3.1: Has the experimental design been preregistered?}}

\noindent \textit{Notes:} If the answer is yes, also give a link to the registration page for the experiment.

\vspace{.1cm}
\noindent \textit{Multiple-choice options (select one)}:
\begin{enumerate}[itemsep=0cm,leftmargin=0.5cm]
	\item[\radiobutton] \egcvalue{Yes (please provide link).} 

	\item[\radiobutton] \egcvalue{No.} 

\end{enumerate}\vspace{0.2cm}

\subsection*{\vspace{-1cm}\qsecboxqtn{Question 3.3.2: By what medium are responses collected?}}

\vspace{-.1cm}
\noindent \textit{What to enter in the text box}:~~Describe the platform or other medium used to collect responses, e.g.\  paper forms, Google forms, SurveyMonkey, Mechanical Turk, CrowdFlower, audio/video recording, etc.

\vspace{.1cm}
\subsection*{3.3.3$\;$~Quality assurance}

\noindent \textit{Notes:} Question 3.3.3.1 records information about the \textit{type(s)} of quality assurance employed, and Question 3.3.3.2 records the details of the corresponding quality assurance methods.

\subsection*{\vspace{-0.5cm}
\qsecboxqtn{Question 3.3.3.1: What types of quality assurance methods are used to ensure that evaluators are sufficently qualified and/or their responses are of sufficient quality?}}

\noindent If any quality assurance methods other than those listed were used, select `other', and describe why below.  If no methods were used, select \textit{none of the above}.

\vspace{.1cm}
\noindent \textit{Check-box options (select all that apply)}:
\begin{enumerate}[itemsep=0cm,leftmargin=0.5cm]
	\item[\checkbox] \egcvalue{Evaluators are required to be native speakers of the language they evaluate}:~~Mechanisms are in place to ensure all participants are native speakers of the language they evaluate. 

	\item[\checkbox] \egcvalue{Automatic quality checking methods are used during and/or after evaluation}:~~Evaluations are checked for quality by automatic scripts during or after evaluations, e.g.\  evaluators are given known bad/good outputs to check that scores are appropriate. 

	\item[\checkbox] \egcvalue{Manual quality checking methods are used during/post evaluation}:~~Evaluations are checked for quality by a manual process during or after evaluations, e.g.\  scores assigned by evaluators are monitored by researchers conducting the experiment. 

	\item[\checkbox] \egcvalue{Evaluators are excluded if they fail quality checks (often or badly enough)}:~~There are conditions under which evaluations produced by participants are not included in the final results due to quality issues. 

	\item[\checkbox] \egcvalue{Some evaluations are excluded because of failed quality checks}:~~There are conditions under which some (but not all) of the evaluations produced by some participants are not included in the final results due to quality issues. 

	\item[\checkbox] \egcvalue{Other (please describe)}:~~Briefly mention any other quality-assurance methods that were used. Details of the method should be entered under 3.3.3.2. 

	\item[\checkbox] \egcvalue{None of the above (no quality assurance methods used).} 

\end{enumerate}\vspace{0.2cm}

\subsection*{\vspace{-1cm}
\qsecboxqtn{Question 3.3.3.2: What methods are used for each of the types of quality assurance methods that were selected in Question 3.3.3.1?}}

\noindent \textit{What to enter in the text box}:~~Give details of the methods used for each of quality assurance types from the last question. E.g.\  if quality checks were used, give details of the check. If no quality assurance methods were used, enter `N/A'.

\vspace{.2cm}
\subsection*{3.3.4$\;$~Form/Interface}
\vspace{.2cm}

\subsection*{\qsecboxqtn{Question 3.3.4.1: Where can the form/interface that was shown to participants be viewed?}}
\vspace{-.1cm}

\noindent \textit{What to enter in the text box}:~~Enter a URL linking to a screenshot or copy of the form if possible. If there are many files, please create a signpost page (e.g.\  on \href{https://github.com}{GitHub}) that contains links to all applicable files.  If there is a separate introductory interface/page, include it under Question 3.2.4.

\subsection*{\vspace{-.6cm}
\qsecboxqtn{Question 3.3.4.2: What types of information are evaluators shown when carrying out evaluations?}}
\vspace{-.1cm}

\noindent \textit{What to enter in the text box}:~~Describe the \textit{types} of information (the evaluation item, a rating instrument, instructions, definitions, etc.) evaluators can see while carrying out each assessment. In particular, explain any variation that cannot be seen from the information linked to in Question 3.3.4.1.

\subsection*{\vspace{-.6cm}
\qsecboxqtn{Question 3.3.5: How free are evaluators regarding when and how quickly to carry out evaluations?}}
\vspace{-.1cm}

\vspace{.1cm}
\noindent \textit{Check-box options (select all that apply)}:
\begin{enumerate}[itemsep=0cm,leftmargin=0.5cm]
	\item[\checkbox] \egcvalue{Evaluators must carry out the evaluation at a specific time/date.} 

	\item[\checkbox] \egcvalue{Evaluators must complete each individual assessment within a set amount of time.} 

	\item[\checkbox] \egcvalue{Evaluators must complete the whole evaluation within a set amount of time.} 

	\item[\checkbox] \egcvalue{Evaluators must complete the whole evaluation in one sitting}:~~Partial progress cannot be saved and the evaluation cannot be returned to on a later occasion. 

	\item[\checkbox] \egcvalue{None of the above (please describe)}:~~Select this option if none of the above are the case in the experiment, then describe any other constraints imposed on when and/or how quickly evaluations must be carried out. 

\end{enumerate}\vspace{0.2cm}

\subsection*{\vspace{-1cm}
\qsecboxqtn{Question 3.3.6: Are evaluators told they can ask questions about the evaluation and/or provide feedback?}}

\vspace{-.1cm}
\noindent \textit{Check-box options (select all that apply)}:
\begin{enumerate}[itemsep=0cm,leftmargin=0.5cm]
	\item[\checkbox] \egcvalue{Evaluators can ask questions during the evaluation}:~~Evaluators are told explicitly that they can ask questions about the evaluation experiment before starting on their assessments, either during or after training. 

	\item[\checkbox] \egcvalue{Evaluators are told they can ask any questions during the evaluation}:~~Evaluators are told explicitly that they can ask questions about the evaluation experiment while carrying out their assessments. 

	\item[\checkbox] \egcvalue{Evaluators provide feedback after the evaluation}:~~Evaluators are explicitly asked to provide feedback and/or comments about the evaluation after completing it, either verbally or in written form, e.g.\  via an exit questionnaire or a comment box. 

	\item[\checkbox] \egcvalue{Other (please describe)}:~~Use this space to describe any other ways you provide for evaluators to ask questions or provide feedback. 

	\item[\checkbox] \egcvalue{None of the above}:~~Select this option if evaluators are not able to ask questions or provide feedback. 

\end{enumerate}\vspace{0.2cm}

\subsection*{\vspace{-1cm}
\qsecboxqtn{Question 3.3.7: What are the conditions in which evaluators carry out the evaluations?}}

\vspace{-.1cm}
\noindent \textit{Multiple-choice options (select one)}:\vspace{-.1cm}

\begin{enumerate}[itemsep=-0.015cm,leftmargin=0.5cm]
	\item[\radiobutton] \egcvalue{Evaluators carry out assessments at a place of their own choosing}:~~Evaluators are given access to the evaluation medium specified in Question 3.3.2, and subsequently choose where to carry out their evaluations. 

	\item[\radiobutton] \egcvalue{Evaluators carry out assessments in a lab, and conditions \underline{are} controlled to be the same for each evaluator.} 

	\item[\radiobutton] \egcvalue{Evaluators carry out assessments in a lab, and conditions \underline{are not} controlled to be the same for different evaluators.} 

	\item[\radiobutton] \egcvalue{Evaluators carry out assessments in a real-life situation, and conditions \underline{are} controlled to be the same for each evaluator}:~~Evaluations are carried out in a real-life situation, i.e.\  one that would occur whether or not the evaluation was carried out (e.g.\  evaluating a dialogue system deployed in a live chat function on a website), and conditions in which evaluations are carried out are controlled to be the same. 

	\item[\radiobutton] \egcvalue{Evaluators carry out assessments in a real-life situation, and conditions \underline{are not} controlled to be the same for different evaluators.} 

	\item[\radiobutton] \egcvalue{Evaluators carry out assessments outside of the lab, in a situation designed to resemble a real-life situation, and conditions \underline{are} controlled to be the same for each evaluator}:~~Evaluations are carried out outside of the lab, in a situation intentionally similar to a real-life situation (but not actually a real-life situation), e.g.\  user-testing a navigation system where the destination is part of the evaluation design, rather than chosen by the user. Conditions in which evaluations are carried out are controlled to be the same. 

	\item[\radiobutton] \egcvalue{Evaluators carry out assessments outside of the lab, in a situation designed to resemble a real-life situation, and conditions \underline{are not} controlled to be the same for different evaluators.} 

	\item[\radiobutton] \egcvalue{Other (please describe)}:~~Use this space to provide additional, or alternative, information about the conditions in which evaluators carry out assessments, not covered by the options above. 

\end{enumerate}\vspace{0.2cm}

\subsection*{\vspace{-1.2cm}
\qsecboxqtn{Question 3.3.8: In what ways do conditions in which evaluators carry out the evaluations vary for different evaluators?}}

\noindent \textit{What to enter in the text box}:~~For those conditions that are not controlled to be the same, describe the variation that can occur.  For conditions that are controlled to be the same, enter `N/A'.

\vspace{0.5cm}
\section*{HEDS Section~4:~Definition and Operationalisation of Quality Criteria}

\noindent \textit{Notes:} Questions in this section record information about each quality criterion (Fluency, Grammaticality, etc.) assessed in the human evaluation experiment that this sheet is being completed for.

If multiple quality criteria are evaluated, the form creates subsections for each criterion headed by the criterion name for each one. These are implemented as overlaid windows with tabs for navigating between them.

\vspace{.1cm}

\subsection*{4.1$\;$~Quality Criterion Properties}

\noindent \textit{Notes:} Questions 4.1.1--4.1.3 capture aspects of quality assessed by a given quality criterion in terms of three orthogonal properties: (i) what type of quality is being assessed; (ii) what aspect of the system output is being assessed; and (iii) whether system outputs are assessed in their own right or with reference to some system-internal or system-external frame of reference. For full explanations see \citet{belz-etal-2020-disentangling}.

\subsection*{\vspace{-.3cm}
\qsecboxqtn{Question 4.1.1: What type of quality is assessed by the quality criterion?}}

\vspace{-.1cm}
\noindent \textit{Multiple-choice options (select one)}:
\begin{enumerate}[itemsep=0cm,leftmargin=0.5cm]
	\item[\radiobutton] \egcvalue{Correctness}:~~Select this option if it is possible to state, generally for all outputs, the conditions under which outputs are maximally correct (hence of maximal quality). E.g.\  for Grammaticality, outputs are (maximally) correct if they contain no grammatical errors; for Semantic Completeness, outputs are correct if they express all the content in the input. 

	\item[\radiobutton] \egcvalue{Goodness}:~~Select this option if, in contrast to correctness criteria, there is no single, general mechanism for deciding when outputs are maximally good, only for deciding for any two outputs which is better and which is worse. E.g.\  for Fluency, even if outputs contain no disfluencies, there may be other ways in which any given output could be more fluent. 

	\item[\radiobutton] \egcvalue{Feature}:~~Select this option if, in terms of property \textit{X} captured by the criterion, outputs are not generally better if they are more \textit{X}, but instead, depending on evaluation context, more \textit{X} may be either better or worse. E.g.\  for Specificity, outputs can be more specific or less specific, but it's not the case that outputs are, in the general case, better when they are more specific. 

\end{enumerate}\vspace{0.2cm}

\subsection*{\vspace{-.9cm}
\qsecboxqtn{Question 4.1.2: Which aspect of system outputs is assessed by the quality criterion?}}

\vspace{-.1cm}
\noindent \textit{Multiple-choice options (select one)}:
\vspace{-.1cm}
\begin{enumerate}[itemsep=0cm,leftmargin=0.5cm]
	\item[\radiobutton] \egcvalue{Form of output}:~~Select this option if the criterion assesses the form of outputs alone, e.g.\  Grammaticality is only about the form, a sentence can be grammatical yet be wrong or nonsensical in terms of content. 

	\item[\radiobutton] \egcvalue{Content of output}:~~Select this option if the criterion assesses the content/meaning of the output alone, e.g.\  Meaning Preservation only assesses content; two sentences can be considered to have the same meaning, but differ in form. 

	\item[\radiobutton] \egcvalue{Both form and content of output}:~~Select this option if the criterion assesses outputs as a whole, not just form or just content. E.g.\  Coherence, Usefulness and Task Completion fall in this category. 

\end{enumerate}\vspace{0.2cm}

\subsection*{\vspace{-0.9cm}
\qsecboxqtn{Question 4.1.3: Is each output assessed for quality in its own right, or with reference to a system-internal or external frame of reference?}}

\vspace{-.1cm}
\noindent \textit{Multiple-choice options (select one)}:
\vspace{-.1cm}
\begin{enumerate}[itemsep=0cm,leftmargin=0.5cm]
	\item[\radiobutton] \egcvalue{Quality of output in its own right}:~~Select this option if output quality is assessed without referring to anything other than the output itself, i.e.\  no system-internal or external frame of reference. E.g.\  Poeticness is assessed by considering (just) the output and how poetic it is. 

	\item[\radiobutton] \egcvalue{Quality of output relative to the input}:~~Select this option if output quality is assessed relative to the input. E.g.\  Answerability is the degree to which the output question can be answered from information in the input. 

	\item[\radiobutton] \egcvalue{Quality of output relative to a system-external frame of reference}:~~Select this option if output quality is assessed with reference to system-external information, such as a knowledge base, a person's individual writing style, or the performance of an embedding system. E.g.\  Factual Accuracy assesses outputs relative to a source of real-world knowledge. 

\end{enumerate}\vspace{0.2cm}

\subsection*{4.2$\;$~Evaluation mode properties}

\noindent \textit{Notes:} Questions 4.2.1--4.2.3 record properties that are orthogonal to quality criterion properties (preceding section), i.e.\  any given quality criterion can in principle be combined with any of the modes (although some combinations are much more common than others).

\subsection*{\vspace{-.6cm}
\qsecboxqtn{Question 4.2.1: Does an individual assessment involve an objective or a subjective judgment?}}

\vspace{-.1cm}
\noindent \textit{Multiple-choice options (select one)}:
\vspace{-.1cm}
\begin{enumerate}[itemsep=0cm,leftmargin=0.5cm]
	\item[\radiobutton] \egcvalue{Objective}:~~Select this option if the evaluation uses objective assessment, e.g.\  any automatically counted or otherwise quantified measurements such as mouse-clicks, occurrences in text, etc. Repeated assessments of the same output with an objective-mode evaluation method should yield the same score/result. 

	\item[\radiobutton] \egcvalue{Subjective}:~~Select this option in all other cases. Subjective assessments involve ratings, opinions and preferences by evaluators. Some criteria lend themselves more readily to subjective assessments, e.g.\  Friendliness of a conversational agent, but an objective measure e.g.\  based on lexical markers is also conceivable. 

\end{enumerate}\vspace{0.2cm}

\subsection*{\vspace{-1cm}
\qsecboxqtn{Question 4.2.2: Are outputs assessed in absolute or relative terms?}}

\vspace{.1cm}
\noindent \textit{Multiple-choice options (select one)}:
\begin{enumerate}[itemsep=0cm,leftmargin=0.5cm]
	\item[\radiobutton] \egcvalue{Absolute}:~~Select this option if evaluators are shown outputs from a single system during each individual assessment. 

	\item[\radiobutton] \egcvalue{Relative}:~~Select this option if evaluators are shown outputs from multiple systems at the same time during assessments, typically ranking or preference-judging them. 

\end{enumerate}\vspace{0.2cm}

\subsection*{\vspace{-1cm}\qsecboxqtn{Question 4.2.3: Is the evaluation intrinsic or extrinsic?}}

\vspace{.1cm}
\noindent \textit{Multiple-choice options (select one)}:
\begin{enumerate}[itemsep=0cm,leftmargin=0.5cm]
	\item[\radiobutton] \egcvalue{Intrinsic}:~~Select this option if quality of outputs is assessed \textit{without} considering their effect on something external to the system such as the performance of an embedding system or of a user at a task. 

	\item[\radiobutton] \egcvalue{Extrinsic}:~~Select this option if quality of outputs \textit{is} assessed in terms of their effect on something external to the system such as the performance of an embedding system or of a user at a task. 

\end{enumerate}\vspace{0.2cm}

\subsection*{4.3$\;$~Response elicitation}

\noindent \textit{Notes:} The questions in this section concern response elicitation, by which we mean how the ratings or other measurements that represent assessments for the quality criterion in question are obtained.  This includes what is presented to evaluators, how they select a response, and via what type of tool, etc.

\subsection*{4.3.1$\;$~Quality criterion name}
\vspace{.1cm}
\subsection*{\qsecboxqtn{Question 4.3.1.1: What do you call the quality criterion in explanations/interfaces to evaluators?}}

\noindent \textit{What to enter in the text box}:~~The name you use to refer to the quality criterion in explanations and/or interfaces created for evaluators. Examples of quality criterion names include Fluency, Clarity, Meaning Preservation. If no name is used, state `no name given'.

\subsection*{\vspace{-1cm}\qsecboxqtn{Question 4.3.1.2: What standardised quality criterion name does the name entered for 4.3.1.1 correspond to?}}

\noindent \textit{What to enter in the text box}:~~Map the quality criterion name used in the evaluation experiment to its equivalent in a standardised set of quality criterion names and definitions such as QCET \citep{belz-etal-2024-qcet-interactive, belz-etal-2025-qcet}, and enter the standardised name and reference to the paper here. In performing this mapping, the information given in Questions 4.3.7 (question/prompt), 3.3.4.1--3.3.4.2 (interface/information shown to evaluators), 4.3.2 (QC definition), 3.2.4 (training/practice), and 4.3.1.1 (verbatim QC name) should be taken into account, in this order of precedence.

\subsection*{\vspace{-0.6cm}\qsecboxqtn{Question 4.3.2: What definition do you give for the quality criterion in explanations/interfaces to evaluators?}}

\noindent \textit{What to enter in the text box}:~~Copy and paste the verbatim definition you give to evaluators to explain the quality criterion they're assessing. If you don't explicitly call it a definition, enter the nearest thing to a definition you give them. If you don't give any definition, state `no definition given'.

\subsection*{\vspace{-0.6cm}\qsecboxqtn{Question 4.3.3: What is the size of the scale or other rating instrument?}}

\noindent \textit{What to enter in the text box}:~~An integer representing the number of different possible response values obtained with the scale or rating instrument. Enter `continuous' if the number of response values is not finite. Enter `N/A' if there is no scale or rating instrument. E.g.\  for a 5-point rating scale, enter `5'; for a slider that can return 100 different values (even if it looks continuous), enter `100'.  If no rating instrument is used (e.g.\  when evaluation gathers post-edits or qualitative feedback only), enter `N/A'.

\subsection*{\vspace{-0.6cm}\qsecboxqtn{Question 4.3.4: What are the possible values of the scale or other rating instrument?}}

\noindent \textit{What to enter in the text box}:~~List, or give the range of, the possible response values returned by the rating instrument. The list or range should be of the size specified in Question 4.3.3. If there are too many to list, use a range. E.g.\  for two-way forced-choice preference judgments collected via a slider, the list entered might be `[-50,+50]'. If no rating instrument is used, enter `N/A'.

\subsection*{\vspace{-.7cm}\qsecboxqtn{Question 4.3.5: How is the scale or other rating instrument presented to evaluators?}}

\noindent \textit{Multiple-choice options (select one)}:
\begin{enumerate}[itemsep=0cm,leftmargin=0.5cm]
	\item[\radiobutton] \egcvalue{Multiple-choice options}:~~Select this option if evaluators select exactly one of multiple options. 

	\item[\radiobutton] \egcvalue{Check-boxes}:~~Select this option if evaluators select any number of options from multiple given options. 

	\item[\radiobutton] \egcvalue{Slider}:~~Select this option if evaluators move a pointer on a slider scale to the position corresponding to their assessment. 

	\item[\radiobutton] \egcvalue{N/A (there is no rating instrument)}:~~Select this option if there is no rating instrument. 

	\item[\radiobutton] \egcvalue{Other (please describe)}:~~Select this option if there is a rating instrument, but none of the above adequately describe the way you present it to evaluators. Use the text box to describe the rating instrument and link to a screenshot. 

\end{enumerate}\vspace{0.1cm}

\subsection*{\vspace{-1cm}\qsecboxqtn{Question 4.3.6: If there is no rating instrument, what is the task the evaluators perform?}}

\noindent \textit{What to enter in the text box}:~~If (and only if) there is no rating instrument, i.e.\  you entered `N/A' for Questions 4.3.3--4.3.5, use this space to describe the task evaluators perform, and what information is recorded. Tasks that don't use rating instruments include ranking multiple outputs, finding information, playing a game, etc.). If there is a rating instrument, enter `N/A'.

\subsection*{\vspace{-.5cm}\qsecboxqtn{Question 4.3.7: What is the verbatim question, prompt or instruction given to evaluators (visible to them during each individual assessment)?}}

\noindent \textit{What to enter in the text box}:~~Copy and paste the verbatim text that evaluators see during each assessment, that is intended to convey the evaluation task to them. E.g.\  \textit{Which of these texts do you prefer?} Or \textit{Make any corrections to this text that you think are necessary in order to improve it to the point where you would be happy to provide it to a client.}

\subsection*{\vspace{-.6cm}\qsecboxqtn{Question 4.3.8: What form of response elicitation is used in collecting assessments from evaluators?}}

\noindent The terms and explanations in this section have been adapted from \citet{howcroft-etal-2020-twenty}.

\vspace{.2cm}
\noindent \textit{Multiple-choice options (select one)}:
\begin{enumerate}[itemsep=0cm,leftmargin=0.5cm]
	\item[\radiobutton] \egcvalue{(Dis)agreement with quality statement}:~~Participants indicate the degree to which they agree with a given quality statement on a rating instrument. The rating instrument is labelled with degrees of agreement and can additionally have numerical labels. E.g.\ \textit{This text is fluent: 1=strongly disagree\ldots5=strongly agree}. 

	\item[\radiobutton] \egcvalue{Direct quality estimation}:~~Participants indicate level of quality on a rating instrument, which typically (but not always) mentions the quality criterion explicitly. E.g.\  \textit{How fluent is this text? 1=not at all fluent\ldots5=very fluent}. 

	\item[\radiobutton] \egcvalue{Relative quality estimation (including ranking)}:~~Participants evaluate two or more items in terms of which is better. E.g.\  \textit{Rank these texts in terms of Fluency}: \textit{Which of these texts is more fluent?} \textit{Which of these items do you prefer?} 

	\item[\radiobutton] \egcvalue{Counting occurrences in text}:~~Evaluators are asked to count how many times some type of phenomenon occurs, e.g.\  the number of facts contained in the output that are inconsistent with the input. 

	\item[\radiobutton] \egcvalue{Qualitative feedback (e.g.\  via comments entered in a text box)}:~~Typically, these are responses to open-ended questions in a survey or interview. 

	\item[\radiobutton] \egcvalue{Evaluation through post-editing/ annotation}:~~Select this option if the evaluators' task consists of editing, or inserting annotations in, text. E.g.\  evaluators may perform error correction and edits are then automatically measured to yield a numerical score. 

	\item[\radiobutton] \egcvalue{Output classification or labelling}:~~Select this option if evaluators assign outputs to categories. E.g.\  \textit{What is the overall sentiment of this piece of text? — Positive/neutral/negative.} 

	\item[\radiobutton] \egcvalue{User-text interaction measurements}:~~Select this option if participants in the evaluation experiment interact with a text in some way, and measurements are taken of their interaction. E.g.\  reading speed, eye movement tracking, comprehension questions, etc. Excludes situations where participants are given a task to solve and their performance is measured which comes under the next option. 

	\item[\radiobutton] \egcvalue{Task performance measurements}:~~Select this option if participants in the evaluation experiment are given a task to perform, and measurements are taken of their performance at the task. E.g.\  task is finding information, and task performance measurement is task completion speed and success rate. 

	\item[\radiobutton] \egcvalue{User-system interaction measurements}:~~Select this option if participants in the evaluation experiment interact with a system in some way, while measurements are taken of their interaction. E.g.\  duration of interaction, hyperlinks followed, number of likes, or completed sales. 

	\item[\radiobutton] \egcvalue{Other (please describe)}:~~Use the text box to describe the form of response elicitation used in assessing the quality criterion if it doesn't fall in any of the above categories. 

\end{enumerate}\vspace{0.2cm}

\subsection*{\vspace{-1cm}\qsecboxqtn{Question 4.3.9: How are raw responses from participants aggregated or otherwise processed to obtain reported scores for this quality criterion?}}

\noindent \textit{What to enter in the text box}:~~Normally a set of separate assessments is collected from evaluators and then converted to the results as reported. Describe here the method(s) used in the conversion(s). E.g.\  macro-averages or micro-averages are computed from numerical scores to provide summarising, per-system results.  If no such method was used, enter `results were not processed or aggregated before being reported'.

\subsection*{\vspace{-1cm}\qsecboxqtn{Question 4.3.10: What method(s) are used for determining effect size and significance of findings for this quality criterion?}}

\noindent \textit{What to enter in the text box}:~~The list of methods used for calculating the effect size and significance of any results, both as reported in the paper given in Question 1.1, for this quality criterion. If none calculated, enter `None'.

\subsection*{4.3.11$\;$~Inter-annotator agreement}

\subsection*{\qsecboxqtn{Question 4.3.11.1: How was the \underline{inter}-annotator agreement between evaluators measured for this quality criterion?}}

\noindent \textit{What to enter in the text box}:~~The method(s) used for measuring \underline{inter}-annotator agreement. If inter-annotator agreement was not measured, enter `InterAA not assessed'.

\subsection*{\vspace{-1cm}\qsecboxqtn{Question 4.3.11.2: What was the \underline{inter}-annotator agreement score?}}

\noindent \textit{What to enter in the text box}:~~The \underline{inter}-annotator agreement score(s) obtained with the method(s) in Question 4.3.11.1.  Enter `InterAA not assessed' if applicable.

\subsection*{4.3.12$\;$~Intra-annotator agreement}

\subsection*{\qsecboxqtn{Question 4.3.12.1: How was the \underline{intra}-annotator agreement between evaluators measured for this quality criterion?}}

\noindent \textit{What to enter in the text box}:~~The method(s) used for measuring \underline{intra}-annotator agreement. If intra-annotateor agreement was not measured, enter `IntraAA not assessed'.

\subsection*{\vspace{-1cm}\qsecboxqtn{Question 4.3.12.2: What was the \underline{intra}-annotator agreement score?}}

\noindent \textit{What to enter in the text box}:~~The \underline{intra}-annotator agreement score(s) obtained with the method(s) in Question 4.3.12.1.  Enter `IntraAA not assessed' if applicable.

\section*{HEDS Section~5:~Ethics}
\vspace{.2cm}
\subsection*{\qsecboxqtn{Question 5.1: Which research ethics committee has approved the evaluation experiment this sheet is being completed for, or the larger study it is part of?}}

\noindent \textit{What to enter in the text box}:~~Normally, research organisations, universities and other higher-education institutions require some form ethical approval before experiments involving human participants, however innocuous, are permitted to proceed. Please provide here the name of the body that approved the experiment, or state `No ethical approval obtained' if applicable.

\subsection*{\vspace{-.5cm}\qsecboxqtn{Question 5.2: Does personal data (as defined in GDPR Art. 4, §1: \href{https://gdpr.eu/article-4-definitions}{https://gdpr.eu/article-4-definitions}) occur in any of the system outputs (or human-authored stand-ins) evaluated, or responses collected, in the experiment this sheet is being completed for?}}

\vspace{.1cm}
\noindent \textit{Multiple-choice options (select one)}:
\begin{enumerate}[itemsep=0cm,leftmargin=0.5cm]
	\item[\radiobutton] \egcvalue{No, personal data as defined by GDPR was neither evaluated nor collected.} 

	\item[\radiobutton] \egcvalue{Yes, personal data as defined by GDPR was evaluated and/or collected}:~~Explain in the text box, how it was ensured that the personal data was handled in accordance with GDPR. 

\end{enumerate}\vspace{0.2cm}

\subsection*{\vspace{-1cm}\qsecboxqtn{Question 5.3: Does special category information (as defined in GDPR Art. 9, §1:  \href{https://gdpr.eu/article-9-processing-special-categories-of-personal-data-prohibited}{https://gdpr.eu/article-9-processing-special-categories-of-personal-data-prohibited}) occur in any of the evaluation items evaluated, or responses collected, in the evaluation experiment this sheet is being completed for?}}

\vspace{.1cm}
\noindent \textit{Multiple-choice options (select one)}:
\begin{enumerate}[itemsep=0cm,leftmargin=0.5cm]
	\item[\radiobutton] \egcvalue{No, special category data as defined by GDPR was neither evaluated nor collected.} 

	\item[\radiobutton] \egcvalue{Yes, special category data as defined by GDPR was evaluated and/or collected}:~~Explain in the text box how it was ensured that the special-category data was handled in accordance with GDPR. 

\end{enumerate}

\subsection*{\vspace{-.8cm}\qsecboxqtn{Question 5.4: Have any impact assessments been carried out for the evaluation experiment, and/or any data collected/evaluated in connection with it?}}

\noindent \textit{What to enter in the text box}:~~If an \textit{ex ante} or \textit{ex post} impact assessment has been carried out, \textit{and} the assessment plan and process, as well as the outcomes, were captured in written form, describe them here and link to the report. Otherwise enter `no impact assessment carried out'. Types of impact assessment include data protection impact assessments, e.g.\  under \href{https://ico.org.uk/for-organisations/uk-gdpr-guidance-and-resources/accountability-and-governance/data-protection-impact-assessments-dpias}{GDPR}.  Environmental and social impact assessment frameworks are also available.

\section{Additional Explanations}\label{sec:add-comments}

\noindent \textit{Meaning of `experiment'}

\vspace{.1cm}

\noindent In the context of HEDS, an experiment consists of a set of assessments for one or more evaluation methods each assessing one quality criterion, that are collected at the same time, with the same experimental design. This means that for a given experiment, all HEDS questions except for those in HEDS Section 4 (about quality criteria) have only one answer. 

\vspace{.3cm}

\noindent \textit{Question 4.3.1.2: What standardised quality criterion name does the name entered for 4.3.1.1 correspond to?}

\vspace{.15cm}

\noindent As discussed in detail elsewhere \cite{belz-etal-2020-disentangling,howcroft-etal-2020-twenty}, just because two evaluation experiments use the same quality criterion name does not mean that they assess the same aspect of quality. The only way we can be sure that the same aspect of quality is being assessed is if we map the two quality criterion names to a single standard set of quality criteria via the same systematic mapping process. 

The QCET taxonomy of quality criteria \cite{belz-etal-2025-qcet} was designed to provide both a standard set of quality criteria names and definitions, and the mapping process.  It does this via the taxonomic structure which is intended to be followed top down on the way to identifying the node that best matches the quality criterion name that is to be standardised.

By using the standardised quality criteria from QCET, one can also identify for each quality criterion, the correct type of quality assessed (Question 4.1.1), aspect of system outputs assessed (Question 4.1.2), and the frame of reference (Question 4.1.3).  These pieces of information are fixed for each quality criterion and can be seen when viewing a quality criterion node in the taxonomy.

\vspace{.2cm}

\appendix

\section*{Appendix}

\section{Changes Since HEDS~2.0}\label{appsec:delta}

\textit{Question numbering:} We have introduced two new questions (4.3.12.1 and 4.3.12.2), and have in seven cases replaced what was a single question in HEDS~2.0 with two or more in 3.0. E.g.\ there was one question on inter-annotator agreement in 2.0 (4.3.11), whereas now there are two (4.3.11.1 and 4.3.11.2). All questions with numbering of depth 4 (e.g.\ 4.3.11.1), and two of depth 3, are the result of such a replacement. In some cases, the motivation was to accommodate a new question without changing other question numbers. In other cases, it was to split an existing question into two for increased clarity and consistency. The complete list of question number mappings from version 2.0 to version 3.0 is as follows:

\noindent \begin{small}
\begin{verbatim}
  1.1 --> 1.1.1, 1.1.2
  1.3 --> 1.3.1.1, 1.3.1.2, 1.3.1.3, 
          1.3.2.1, 1.3.2.2, 1.3.2.3
  3.1.3 --> 3.1.3.1, 3.1.3.2, 3.1.3.3
  3.2.2 --> 3.2.2.1, 3.2.2.2, 3.2.2.3, 
            3.2.2.4
  3.3.3 --> 3.3.3.1, 3.3.3.2
  3.3.4 --> 3.3.4.1, 3.3.4.2
  4.3.11 --> 4.3.11.1, 4.3.11.2
  + 4.3.12.1, 4.3.12.2
\end{verbatim}
\end{small}

In all other cases, questions are in essence the same (apart from rewording), and have the same number, in both versions.

\textit{Question wording:} Most questions have undergone some degree of rewording in order to make them (a) clearer and easier to answer, and (b) more consistent in working and style. 

\textit{Answer types:} In a small number of cases we have replaced a text box answer with a list of options, to achieve greater comparability in answers between users.

The overall motivation for all changes was to make it easier for users to complete the datasheet consistently and comparably (to other users).

\bibliography{main}
\bibliographystyle{acl_natbib}

\end{document}